 \pdfoutput=1 
\documentclass{egpubl}
\BibtexOrBiblatex
\usepackage[backend=biber, bibstyle=EG,citestyle=alphabetic]{biblatex} 
\usepackage{eurovis2023}
\usepackage[outdir=./]{epstopdf}

\SpecialIssuePaper         

\CGFStandardLicense

\usepackage[T1]{fontenc}
\usepackage{dfadobe}  
\usepackage{colortbl}
\usepackage{enumitem}
\usepackage{xcolor}
\definecolor{lightgray}{RGB}{212,212,212}
\usepackage{multirow}
\usepackage{tabu}                      
\usepackage{booktabs}                  
\usepackage{subcaption}

\electronicVersion
\PrintedOrElectronic
\ifpdf \usepackage[pdftex]{graphicx} \pdfcompresslevel=9
\else \usepackage[dvips]{graphicx} \fi

\usepackage{egweblnk}
\usepackage{amssymb}
\usepackage{pifont}

\newcommand{\revise}[1]{#1}

\newcommand{\bpstart}[1]{\vspace{1mm}\noindent{\textbf{#1.}}}

\usepackage{cleveref}

\title[Sequence Summarization Comparison]%
      {A Comparative Evaluation of Visual Summarization Techniques\\for Event Sequences}

\author[Kazi Tasnim Zinat, Jinhua Yang, Arjun Gandhi, Nistha Mitra \& Zhicheng Liu]
{\parbox{\textwidth}{\centering{Kazi Tasnim Zinat$^{1}$\orcid{0000-0001-7914-5955}, Jinhua Yang$^{1}$, Arjun Gandhi$^{1}$, Nistha Mitra$^{1}$ \& Zhicheng Liu$^{1}$\orcid{0000-0002-1015-2759}}
}
\\
{\parbox{\textwidth}{\centering $^1$ University of Maryland College Park, Maryland, United States}}
}

\begin{document}
\vspace{-2mm}
\teaser{\includegraphics[width=\textwidth]{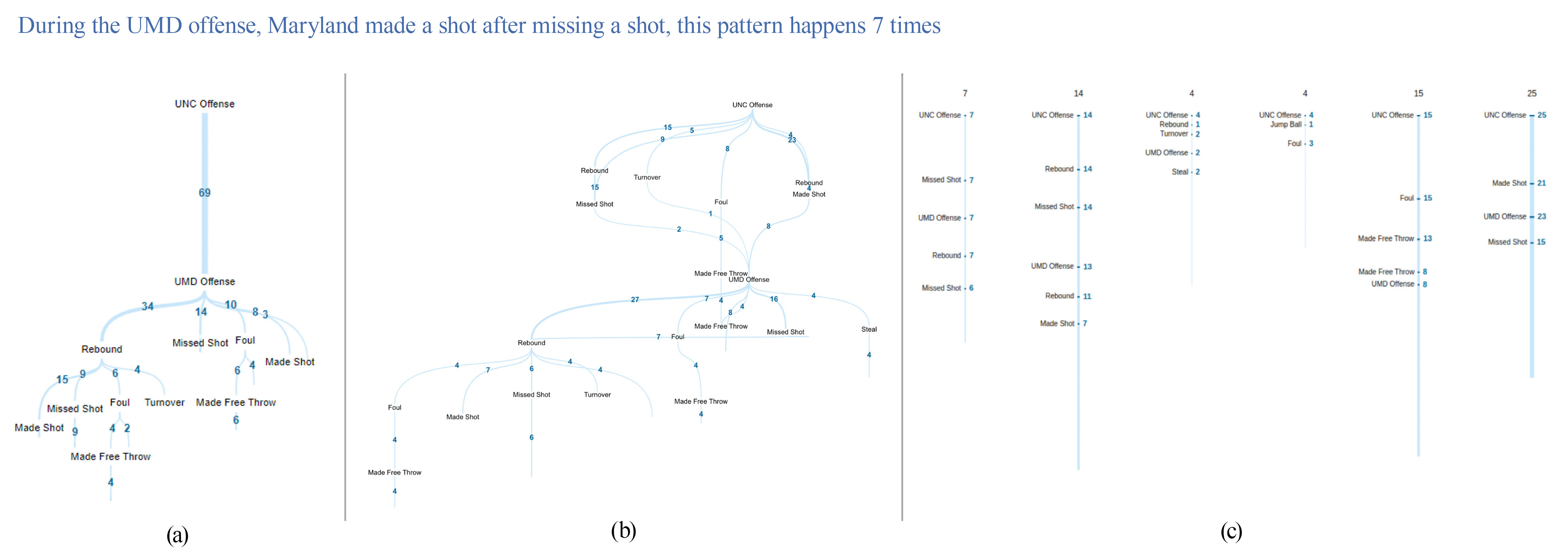}
 \centering
  \caption{In our study, we generated visual summaries of the same dataset using three different techniques: (a) CoreFlow \cite{Liu_2017}, (b) SentenTree \cite{Hu_2017}, and (c) Sequence Synopsis \cite{Chen_2018}. These visual summaries were shown one at a time, and the participants were asked to rate how closely the visualization depicted given insights about the dataset and provide a justification for their ratings.\\}
\label{fig:teaser}
\setlength{\abovecaptionskip}{-3pt}
\setlength{\belowcaptionskip}{-13pt}
}

\maketitle
\begin{abstract}
Real-world event sequences are often complex and heterogeneous, making it difficult to create meaningful visualizations using simple data aggregation and visual encoding techniques. Consequently, visualization researchers have developed numerous visual summarization techniques to generate concise overviews of sequential data. These techniques vary widely in terms of summary structures and contents, and currently there is a knowledge gap in  understanding the effectiveness of these techniques. In this work, we present the design and results of an insight-based crowdsourcing experiment evaluating three existing visual summarization techniques: CoreFlow, SentenTree, and Sequence Synopsis. We compare the visual summaries generated by these techniques across three tasks, on six datasets, at six levels of granularity. We analyze the effects of these variables on summary quality as rated by participants and completion time of the experiment tasks. Our analysis shows that Sequence Synopsis produces the highest-quality visual summaries \revise{for all three tasks}, but understanding Sequence Synopsis results also takes the longest time. We also find that the participants evaluate visual summary quality based on two aspects: content and interpretability. We discuss the implications of our findings on developing and evaluating new visual summarization techniques.


\end{abstract}
\vspace{-5mm}
\section{Introduction}
In many application domains, discrete events are recorded for specific entities, and ordered temporally to form sequences. For example, healthcare providers keep records of lab results or treatment
events for each patient; businesses collect clickstreams to increase conversion; software developers log user behavior to identify
potential usability issues. These datasets are often complex and heterogeneous: few sequences are identical to each other, and there is usually high variability
between sequences in terms of the number and type of events and their orders. Visualizations based on simple visual encoding and aggregation are therefore inadequate.  Extensive research has thus focused on techniques that combine computational methods with visual interfaces \cite{Wei_2012, Monroe_2012, Gotz_2013,Guo_2018,Guo_2019,Magallanes_2021}. In particular, a number of techniques try to generate \textit{visual summaries} of event sequences \cite{Liu_2017_1,Liu_2017,Chen_2018,Chen_2018_StageMap,Monroe_2013, Perer_2014}
by showing only important
events and salient structures that serve as overviews. Figure \ref{fig:teaser} shows exemplary visual summaries of a basketball match dataset consisting of $69$ sequences and $465$ events, generated by three techniques. 

Despite advances in novel visual summarization techniques, we have little understanding of their
effectiveness. To date, there have been no empirical studies comparing these techniques through controlled experiments. The lack of systematic evaluation is problematic: researchers have no established baselines and methods to measure and innovate new techniques; practitioners have no guidance on choosing a suitable technique for their data and analytic needs. 

In this paper, we present the design and results of an insight-based crowdsourcing experiment evaluating three existing visual summarization techniques: CoreFlow, SentenTree, and Sequence Synopsis. We chose these techniques based on considerations such as applicability across different domains, diversity of summary structures, and adjustable summary granularity. \revise{Our focus is on the underlying algorithms, not interactive systems. The algorithms enable \textit{automated} generation of visual summaries, which serve as overviews in visualization
\cite{Shneiderman_1996, Keim_2008}.}  


In the experiment, the participants evaluate how closely visual summaries generated by the techniques at different granularity levels match known ground truths
for different datasets. We analyze the participants' ratings, the time spent on evaluating the summaries, as well as their justifications for the ratings. We find that (1) visual summaries generated by Sequence Synopsis receive the highest ratings, but they also require more time to understand; (2) two factors influence the perceived quality of a visual summary: content and interpretability.  We discuss the implications of our findings on developing and evaluating new summarization techniques.

\noindent This paper makes the following contributions:
\begin{itemize}[noitemsep,topsep=-2pt]
    \item \bpstart{Experiment Design} To the best of our knowledge, this is the first controlled experiment to compare event sequence visual summarization techniques. We identify dataset, task, granularity as independent variables and measure technique effectiveness through user rating, completion time, and text responses.
    \item \bpstart{Result Analysis} Our analysis of the experiment data deepens the understanding on factors influencing summary effectiveness, criteria for assessing summary of quality, and trade-offs in event sequence visual summarization. 
    \item \bpstart{System Implementation} We re-implemented three existing automated event sequence summary techniques as well as three different approaches to visualize summary structures. We plan to open source these implementations. 
\end{itemize}
 
 \vspace{-2mm}

\section{Background and Related Work}
\label{sec:rw}




\subsection{Visualization Techniques for Event Sequence Data}
\begin{figure}[ht]
\centering
  \includegraphics[width=0.4\textwidth]{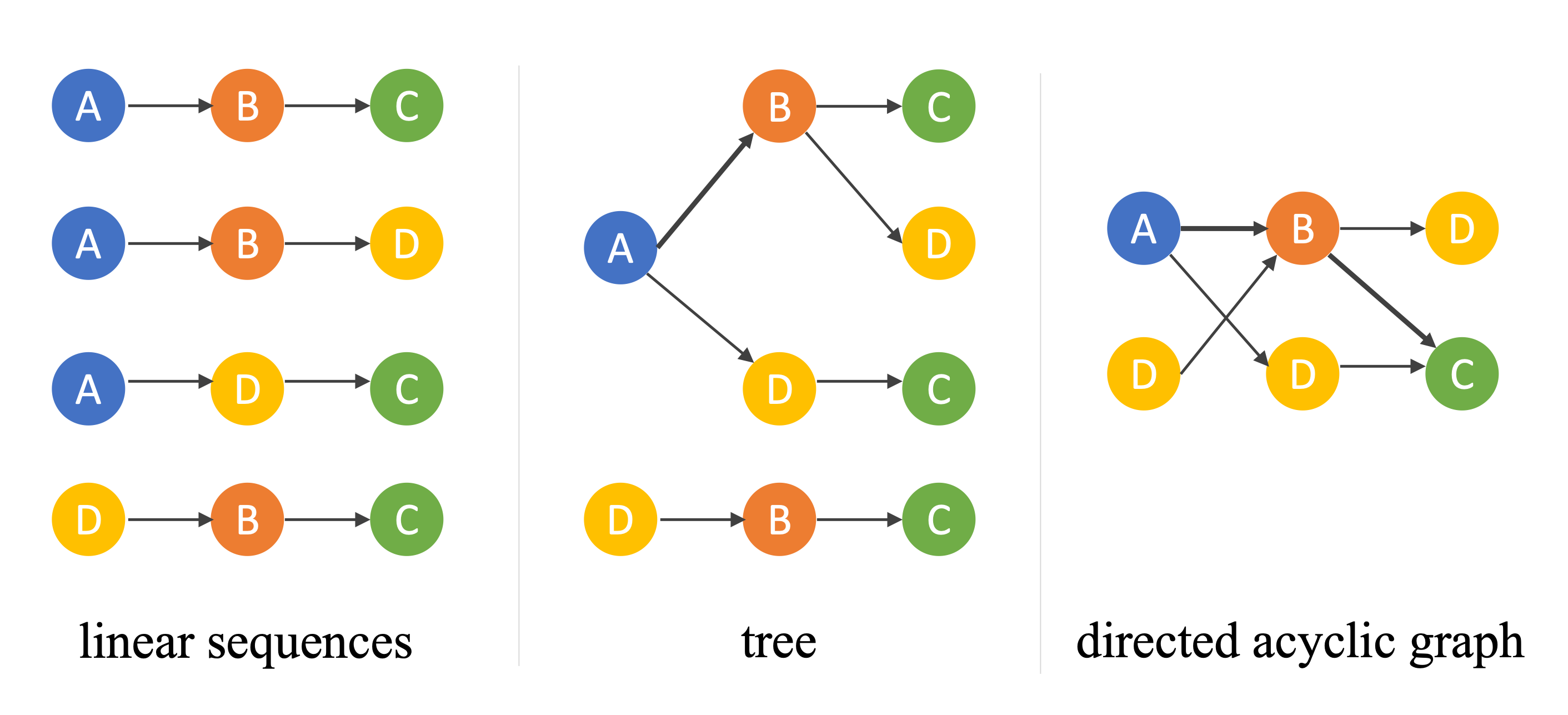}
   \setlength{\belowcaptionskip}{-8pt}
    \setlength{\abovecaptionskip}{0pt}
  \caption{Three main approaches to visualize event sequences as linear sequences, a tree
with a virtual root, and a directed acyclic graph. Each circle
represents an event, order goes from left to right.}
  \label{fig:aggregation}
\end{figure}

\vspace{-1mm}
Early event sequence visualization tools focused on displaying each individual records, \cite{Karam_1994, Plaisant_1996, Plaisant_1998, Harrison_1994,Wang_2008,Fails_2006}. 
and they can only handle a small number of sequences.
Later tools aggregate events across sequences to generate tree structures or directed-acyclic graph (DAG) structures \cite{ Wongsuphasawat_2011, Wongsuphasawat_2012, Monroe_2012, Monroe_2013, Gotz_2013, Perer_2013, Gotz_2014}. Figure \ref{fig:aggregation} shows these aggregation approaches.
A tree structure can then be visualized as an icicle plot \cite{ Wongsuphasawat_2011} or a sunburst chart \cite{Stasko_2000}, and a graph structure can be visualized using a node-link diagram, a Sankey diagram \cite{Wongsuphasawat_2012, Gotz_2013, Gotz_2014}, or concatenated adjacency matrices \cite{Zhao_2015}.
 Besides aggregation, these tools also support additional functionalities, such as querying and filtering to simplify visual overviews of complex sequences
\cite{Monroe_2012, Vrotsou_2009}. 
Many of these functionalities require human knowledge  to interactively generate a meaningful visualization.


\vspace{-2mm}
\subsection{Visual Summarization: Mining and Visualizing Patterns}
\vspace{-1mm}
\revise{It is time consuming to interactively aggregate and simplify events, and this method is often} not scalable enough for existing datasets \cite{Liu_2017_1,Monroe_2013, liu2016mining}. 
Automated techniques have been thus developed to extract important events and patterns from a dataset, and display the extracted results in a concise visual summary. Most automated techniques mine frequent patterns as part of the summarization process, and the structures of the extracted patterns are also in the form of linear sequences, tree, or directed acyclic graphs. For example, Frequence \cite{Perer_2014}, \revise{Patterns and Sequences} \cite{Liu_2017_1}, Chronodes \cite{Polack_2018}, and Peekquence \cite{kwon_peekquence_2016} extract linear sequential patterns. 
Sequence Synopsis \cite{Chen_2018} performs sequence clustering first, and then mines frequent sequential patterns based on an information-theoretic approach to minimize description length \cite{Grunwald_2007}. Instead of extracting linear patterns, 
CoreFlow \cite{Liu_2017} mines branching patterns using a 
recursive rank-divide-trim approach. Despite the misleading term ``tree'' in its name, SentenTree \cite{Hu_2017} uses a breadth first approach to find graph-like patterns in tweets.
 This technique can be applied to event sequence data in general. The extracted patterns can then be visualized according to their structures. Sequence Synopsis \cite{Chen_2018} and \revise{Patterns and Sequences}  \cite{Liu_2017_1} display the sequential patterns as linear ordered event sets. CoreFlow \cite{Liu_2017} visualizes the extracting branching pattern using an icicle plot. SentenTree \cite{Hu_2017} uses a node-link diagram to show the extracted DAG pattern. Linear patterns can also be merged and visualized as a Sankey diagram, as shown in Frequence \cite{Perer_2014}. 

\vspace{-3mm}

\subsection{Segment, Align, and Cluster Event Sequences}
\vspace{-1mm}

Besides mining-based approaches, researchers have also developed visual analytic techniques to segment, align, and cluster sequences. 
To segment sequences, EventThread \cite{Guo_2018} and EventThread2 \cite{Guo_2019} uses the tensor analysis with an unsupervised stage analysis algorithm to find progression states in event sequences.  STBins ~\cite{Qi_2020} renders segment similarity based on temporal binning with Jaccard coefficient-based segment similarity measure. DPvis \cite{Kwon_2020} encodes event sequences via hidden Markov model to identify disease progression pathways. To align sequences by key events, Sequence Braiding \cite{Bartolomeo_2020} computes pairwise alignment of input sequences, and orders them through a constrained intersection reduction algorithm. 
To cluster sequences, Wei et al. \cite{Wei_2012} clusters sequences using a self-organizing map. Cadence \cite{Gotz_2020} supports dynamic hierarchical aggregation of high dimensional event sequence data. Sequen-C \cite{Magallanes_2021} tries to combine both sequence clustering and alignment to support multi-level visual analytics. All these techniques still try to show all the events and sequences with additional visual structures such as alignment and clusters. In this paper, our focus is on the evaluation of mining-based techniques that have a data-reduction component, where the analytical results contain much fewer events and sequences than the original dataset, and the visualizations are best considered as \textit{visual summaries} of event sequences.

\vspace{-2mm}
\subsection{Interactive Visual Analytics}
\vspace{-1mm}
Most event sequence visual analytics works provide interfaces to perform interactive analysis. For example, EventPad  \cite{Cappers_2018}, MAQUI  \cite{Law_2019}, DecisionFlow  \cite{Gotz_2014}, and s|queries \cite{Zgraggen_2015} allow users to query events and patterns. In Eloquence \cite{Vrotsou_2019}- users can interactively add local constraints while patterns are mined using PrefixSpan \cite{Pei_2004}.  Segmentifier \cite{Dextras-Romagnino_2019} proposes a high-level analysis model with support for data wrangling and downstream data analysis. Progressive Insights  \cite{Stolper_2014} provide intermediate results by modifying BFS SPAM mining to prioritize interesting patterns and prune uninteresting ones. Whether automatic summarization techniques are used without human input or in a mixed-initiative approach, their effectiveness plays an important role in determining the analytic outcomes. 
Our goal in this paper is to evaluate the effectiveness of automated summarization techniques for providing such overviews.
\vspace{-2mm}

\section{Motivation and Approach}
\revise{We surveyed $14$ research papers that proposed novel visual summarization techniques over the past $8$ years \cite{Chen_2018, Chen_2018_StageMap, Guo_2018, Hu_2017, kwon_peekquence_2016, Liu_2017, Liu_2017_1, Perer_2014, Polack_2018, Perer_2015, Guo_2019, Gotz_2014_1, Robinson_2017, wang2016unsupervised}. One
paper \cite{Liu_2017} explored comparison with other techniques, but only showed sample visualizations for illustrative purposes. $13$ out of $14$ evaluated the proposed techniques through
qualitative case studies with domain experts. While case studies demonstrate the ecological
validity of the work, they do not provide an objective account on how the techniques compare to each other.} 

\revise{In this paper, we use the term \textit{techniques} to refer to the underlying \textit{algorithms}, not the interactive systems, in the original papers. The algorithms determine both the summary \textit{content} (i.e. what events and relations are included) and summary \textit{structure} (i.e. ~linear sequences, tree, DAG). We focus on evaluating algorithms  based on the following reasons: 1) the algorithms are usually presented as primary contributions in the original papers, 2) the case studies in these papers show that users rely on the algorithmically generated visual summaries as overviews to understand their datasets, and 3) the different user interface and interaction designs in the original papers introduce confounding factors, making it hard to understand technique effectiveness.} 

\revise{To date, there has been no established methods or metrics to evaluate event sequence summarization techniques. Existing studies on visualization effectiveness usually focus on graphical perception  \cite{Heer_2010, Kim_2018}, where the methods and results cannot be directly applied to our problem. First, these studies 
exclusively focus on the choice of encoding methods (i.e., how data attributes are represented
using visual properties of marks). In our case, however, significant data
reduction is performed before visual encoding to make the visualizations readable. Merely focusing on encoding overlooks the importance of data reduction. Second, most of the prior studies
focus on low-level tasks such as looking up and comparing data values. Understanding how well
users perform these tasks does not shed light on the more important questions on the quality of visual summaries. For example, while we care if users can read and understand the visual summary
presented to them,
we are also interested in assessing the ease or difficulty to detect the presence of patterns in the visual summary.
}

\revise{Given the lack of established methods, we have considered different experimental settings. First, we may simply evaluate the algorithms by objectively checking if the generated visual summaries contain pre-formulated ground truth associated with a dataset. However, this approach overlooks the importance of graphical perception: people's ability to identify the ground truth can be influenced by both the summary structure and content. An in-lab controlled study can address this problem. However,  it is important to include multiple datasets from different domains in the study since dataset domain and properties can influence technique effectiveness \cite{Liu_2017}. Furthermore, we want to compare at least three techniques, and include multiple levels of summary granularity. An in-lab design is not likely to scale well for these factors.} 

\revise{Based on these considerations, we propose to adapt the insight-based methodology \cite{Saraiya_2005, North_2006} in a crowdsourcing experiment. We focus on how well the visual summaries  generated by different techniques depict pre-formulated insights about existing datasets, as judged by the participants. An alternative approach where users explore the generated visual summaries to reach insights would be more ecologically valid, but such open-ended designs are difficult to control and monitor on a crowdsourcing platform.  We argue it is a reasonable proxy to test if a visual summary depicts given insights: it would be harder to reach insights if the summary presents less relevant information or is hard to interpret.} 



\section{Insight-Based Crowdsourcing Experiment Design}

\subsection{Visual Summarization Techniques}
\vspace{-1mm}


Dozens of visual summarization techniques for event sequences are available \cite{Guo_2020}. We choose the techniques to be included in this study based on the following criteria. First, the techniques should be domain agnostic and can be applied to datasets from different problem domains. Second, \revise{we want to include three types of summary structures: linear sequences, tree, and directed acyclic graph (DAG).}
Third, the techniques should generate summaries consisting of much fewer events and sequences compared to the input dataset; simple clustering methods, for example, do not satisfy this requirement. Fourth,  we focus on automated summarization techniques in this study, so the generation of visual summaries should require minimal human input. Finally, we would like to compare the visual summaries at different levels of granularity, so the techniques should support controlling summary granularity through appropriate parameters. 

Based on these criteria, we choose CoreFlow \cite{Liu_2017}, SentenTree \cite{Hu_2017}, and Sequence Synopsis \cite{Chen_2018}. These techniques mine frequent events and patterns and visualize them as linear sequences, a tree, and a directed acyclic graph, respectively. \revise{SentenTree \cite{Hu_2017} is the only technique we know that produces a DAG summary. Although it was originally developed for text data, it can be directly extended to event sequences.} They all have granularity parameters that can be tuned, and can be applied to event sequences from different domains. 
\vspace{-2mm}


\subsubsection{Overview of Algorithmic Approaches}
CoreFlow \cite{Liu_2017} recursively applies a rank-divide-trim approach. Events are initially ranked using a pre-defined metric, such as the frequency of occurrence and average index \revise{(the mean value of index positions)} across sequences. The top-ranked event is added to the summary;  and the sequences are partitioned into two groups based on whether they contain the top-ranked event. Finally, the sequences containing the top-ranked event are trimmed. These three operations are recursively applied to resulting sequence groups until either all sequences have been processed or a predefined minimum support threshold is reached.
(i.e., a threshold below which the mining algorithm will stop). 

  SentenTree \cite{Hu_2017} 
  also uses a rank and divide approach. But instead of pruning the sub-sequences up till the first occurrence of the top ranked event, SentenTree also extracts frequent patterns above the minimum support in these sub-sequences. Given the same minimum support, SentenTree usually mines more events and patterns than CoreFlow. 

Sequence Synopsis \cite{Chen_2018} uses the minimum description length \cite{Grunwald_2007} principle to cluster sequences and identify a representative sequential pattern per cluster. The algorithm performs iterative merging to find clusters and associated patterns, while optimizing for the number of generated patterns, and the edits required to obtain the original dataset from the patterns.
\vspace{-2mm}



\subsubsection{Visualization Design}


\revise{Our goal is to compare the summarizing algorithms. However, the visual representations and styles for the generated summaries in the original works vary greatly. To eliminate the potential confounding effects, we did not follow the original visualization designs. Instead, 
we chose a minimalistic and consistent design for the visualizations (Figure \ref{fig:teaser}).}
Each event is represented by a node with a label, events connected by links form a pattern. The vertical position represents the order of events, going from top to bottom. 
The width of a link is proportional to the number of sequences.
To facilitate reading, we place numeric labels representing the number of sequences next to each node in Sequence Synopsis, and on top of each link in CoreFlow and SetenTree. 


We use the tidy tree layout \cite{Reingold1981TidierDO}  for CoreFlow and the Sugiyama layout \cite{Sugiyama_1981} for SentenTree. \revise{The tidy tree layout tries to achieve symmetry and compactness in node positioning.
The Sugiyama layout layers nodes and optimizes their placement and ordering to reduce edge crossings.}
For Sequence Synopsis, we set the patterns in an equidistant layout. 
Vertical position of the events encode the average index position across sequences they appear in. The start and end nodes of mining are hidden to reduce visual clutter. In SentenTree, if a node has two or more predecessors, then the average position is set after the predecessor with the highest average index position. This design decision guarantees a node will always appear after all its predecessors. Finally, the same color scheme is used across techniques for consistency.
\vspace{-2mm}





\subsubsection{Implementation}
\label{sec: implementation}
Among these three techniques, only SentenTree \cite{Hu_2017} has been open-sourced. 
The available implementation assumes tweets as input, and cannot readily handle generic event sequence data. The visualization implementation does not conform to the design guidelines discussed above either. Therefore, we re-implemented the summarization algorithms 
\href{https://osf.io/ha7gm/?view\_only=63a77bb0478f4b5c84b695dddee0bb41}{The supplemental materials}
 contain descriptions of our implementation and how we verify its correctness.
To avoid computational latency being a confounding factor, we pre-compute the summaries, 
render visualizations based on the results, and save the visualization as images to be used in the study.

\subsubsection{Granularity}
Each of these techniques supports parameterized tuning of summary granularity. CoreFlow and SentenTree use minimum support to determine the percentage of sequences an event must appear in to be mined and included in the pattern. Sequence Synopsis uses two parameters $\alpha$ and $\lambda$. $\alpha$  balances the trade-off between minimizing information loss and reducing visual clutter. $\lambda$  controls the total number of patterns. We select to control $\lambda$ as the granularity parameter. \revise{In order to find the right balance between visual clutter and missing information, while maintaining a wide range of variation, we experimented with varying degrees of granularity.}
We \revise{decided upon using} six granularity levels for each technique, with the minimum support value ranging from $5$\% to $30$\%  with increments of $5$\% for CoreFlow and SentenTree, and value of $\lambda$ ranging from $90\%$ to $15\%$ with decrements of 15\% for Sequence Synopsis. 
\vspace{-3mm}
\begin{table*}[h]\fontfamily{ptm}\selectfont\small
\centering

{\small \begin{tabular}{| p{0.15\linewidth}| p{0.35\linewidth} | p{0.065\linewidth} | p{0.03\linewidth} | p{0.03\linewidth} | p{0.045\linewidth} |p{0.06\linewidth} |p{0.05\linewidth} |}
\arrayrulecolor{lightgray}
\hline		
\rowcolor[gray]{.95}  
\multicolumn{1}{|p{0.15\linewidth}|}{\textbf{Dataset}} &
\multicolumn{1}{|p{0.35\linewidth}|}{\textbf{Description}} &
\multicolumn{1}{|p{0.065\linewidth}|}{\textbf{\# Unique }} &
\multicolumn{3}{|p{0.155\linewidth}|}{\textbf{Sequence Length}} & 
\multicolumn{1}{|p{0.06\linewidth}|}{\textbf{\# }} &
\multicolumn{1}{|p{0.05\linewidth}|}{\textbf{Total}}\\ 
\cline{4-6}
\rowcolor[gray]{.95}  
\multicolumn{1}{|c|}{} &
\multicolumn{1}{|c|}{} &
\multicolumn{1}{|p{0.065\linewidth}|}{\textbf{Events}} &
\multicolumn{1}{|p{0.03\linewidth}|}{\textbf{Min}} &
\multicolumn{1}{|p{0.03\linewidth}|}{\textbf{Max}} &
\multicolumn{1}{|p{0.045\linewidth}|}{\textbf{Median}} &
\multicolumn{1}{|p{0.06\linewidth}|}{\textbf{Sequences}} &
\multicolumn{1}{|p{0.05\linewidth}|}{\textbf{Events}}\\
  \hline  
Pediatric Trauma Unit (Trauma)	\cite{TraumaPaper, TraumaDemo} &	Order of trauma response events for children		& 11	& 5 & 11 & 9		& 215 	& 1,991		\\ \hline	

Emergency Department (Emergency) \cite{eventFlowDemo}	&	Patients’ movement through the hospital after being brought to emergency		& 6		& 3 & 16 & 4.5	& 100 & 451 \\ \hline				

UMD vs UNC Basketball Match (Basketball) \cite{BasketballDemo}	&	A play-by-play event log of a basketball match		& 13		& 4 & 13 & 6	& 69 & 465  \\ \hline	

VAST Challenge (VAST) \cite{VAST2017}	&	Published as a challenge in 2017 VAST. The dataset is about the movement of cars in a
nature preserve. Each sequence records the locations a car passed by during its trip		& 6		& 2 & 49 & 8		& 1,000 & 9,443  \\ \hline	

Issue Workflow \cite{WorkflowDataset}  &	Workflows related to bug fixes on an Apache software project	& 16		& 2 & 21 & 11		& 45 & 177 \\ \hline

Career \cite{mauriello2016simplifying, Gu02022Interpretable, Guo_2018, Guo_2019}	&	Career path milestone events for university professors over 23 years	& 10		& 11 & 32 & 17		& 40 & 767 \\ \hline

\end{tabular}}
\caption{Dataset description and event details}
  \label{tab:dataset}
\end{table*}

\begin{table*}[h]\fontfamily{ptm}\selectfont\small
\centering

{\small \begin{tabular}{| p{0.07\linewidth}| p{0.06\linewidth} | p{0.13\linewidth} | p{0.6\linewidth} |}
\arrayrulecolor{lightgray}
\hline		
\rowcolor[gray]{.95}  \textbf{Dataset} & \textbf{Domain} & \textbf{Task} & \textbf{Example Insight}  \\  \hline  

Trauma	&	Medical		& Anomaly Detection	& Only about half out of 215 patients went through the process in the right order: airway→ breathing → pulse → gcs → secondary\_survey		\\ \hline	

Emergency	&	Medical		& Common Pattern Identification	& Out of the 100 people, about a third (37 people) are discharged alive after going to the emergency room	\\ \hline

Basketball	&	Sports		& Common Pattern Identification	& During the UMD offense, Maryland made a shot after missing a shot, this pattern happens 7 times	\\ \hline

VAST	&	Transport		& Clustering	& Approximately 17 cars (17\%) are “pass-throughs"	\\ \hline

Workflow	&	Technical		& Clustering	&  For 6 issues that were created, nothing happened afterwards.	\\ \hline

Career	&	Academic		& Anomaly Detection	&  8 persons do not have any publications or attend any conferences after they became a professor	\\ \hline

\end{tabular}}
\caption{\revise{Domain and assigned task information of the datasets along with example}  insight} \label{tab:task}
\vspace{-2mm}
\end{table*}

\subsection{Datasets, Analytical Tasks and Insights}
\vspace{-1mm}
We searched for event sequence datasets by reviewing published papers at visualization and mining conferences and journals, and  examining public dataset repositories \cite{UCLA_ML, Event_event}. We collected a candidate set of $15$ potential event sequence datasets from various domains. \revise{To reduce bias, we exclude datasets used in any of the three original papers.} We identified the following information:


\begin{itemize}
    \item Application Domain: Example domains include but are not limited to healthcare, software, sports, and human activities. 
    \item Data Size: We record statistics about the size of the dataset in terms of the number of unique events, number of total events, number of sequences, and max/min/median sequence length.
    \item Insights: \revise{There is no established benchmark ground truth readily available for our intended evaluation task. However,} many datasets have associated insights that can be used as ground truth. These insights are usually included in the case study section of corresponding publication; sometimes supplemental videos also present these insights in detail.
    For each dataset, we curated a set of insights both from the paper and the video transcript (if available). Table \ref{tab:task} shows example insights. 
    \item Analytical Tasks: After curating the insights for each dataset, we identify the representative high-level analytical tasks that these insights support. We identified three tasks: Anomaly Detection, Common Pattern Identification, and Clustering.
\end{itemize}
Based on these dimensions, we select six datasets to be used in the study. The goal is to include diverse application domains and varying data size, and to have the same number of datasets for each of the three analytical tasks. \revise{
As we prioritized datasets from different domains with insights from previous studies, we have limited control over different size parameters such as number of unique events and sequence length.} 
We use the associated insights in the corresponding publications and supplemental videos for Emergency \cite{eventFlowDemo}, Trauma \cite{TraumaPaper, TraumaDemo}, Career \cite{mauriello2016simplifying, Gu02022Interpretable, Guo_2018, Guo_2019} and Basketball \cite{BasketballDemo} dataset. For the VAST \cite{VAST2017} and Workflow \cite{WorkflowDataset} datasets, 
 two of the authors independently analyzed these using EventFlow \cite{Monroe_2013} for the Clustering task, and calibrated the findings. We then picked one primary analytical task for each dataset, and identified three insights that can support the task. Table \ref{tab:dataset} contains a brief description of the selected datasets and complexity.
Table \ref{tab:task} has the domain information and task for each dataset with an example insight.

\vspace{-3mm}
\subsection{Study Design}
\label{sec:study_design}
\vspace{-1mm}
\revise{We used an insight-based approach to design our experiment, showing participants one visual summary at a time from a dataset and asking them to rate how accurately the summary matches each of the three insights associated with that dataset. With three techniques, six granularity levels, three analytical tasks, and six datasets (two per task), we generated $3\times$6$\times$3$\times$2 = 108 unique visual summaries. With three (3) insights for each dataset, we have a total of 108$\times$3 = 324 unique combinations.}
It is not possible for a single participant to experience all these combinations in a reasonable amount of time, \revise{since we have to explain the meaning of each dataset to them. We considered various plausible combinations. Finally, we decided that it was important to expose every participant to all three techniques, without overwhelming them with the unfamiliar events of multiple datasets, while also avoiding learning effect. Therefore, we} choose a mixed approach using a within-participants design for technique and insight, and a between-participants design for dataset, task, and granularity. That is, each participant would experience 9 combinations in total: all three (3) techniques and all three (3) insights associated with a dataset, and only one (1) granularity level and one (1) dataset, which corresponds to one (1) task.  
We implemented the study design in JavaScript on Qualtrics 
to  
\revise{guarantee equal participant distribution across granularity levels and datasets, and to ensure participants accurately encounter the combinations mentioned above.}


\vspace{-4mm}

\subsection{Study Procedure}
\vspace{-1mm}

\revise{We developed a three phase evaluation process to ensure the participants have adequate visual and data literacy.}
The participants first complete a tutorial and a pre-screening test. They proceed to the main experiment if they satisfy the prescreening criterion.
\revise{Details of participant selection criteria is mentioned in \ref{subsec: participants}.}
 
\vspace{-2mm} 
\subsubsection{Tutorial}
\vspace{-1.5mm} 

We generated visual summaries using the three techniques on check-in information of $25$ individuals randomly sampled  from the New York City Foursquare check-in dataset \cite{Yang_2015_Foursquare}. The $20$-month check-in data was divided into weeks. This dataset is not part of the main experiment.  
In the tutorial, the participants click through a series of displays, which incrementally highlights different parts of the visual summaries to describe what the nodes and links signifies. The tutorial in particular focuses on the interpretation of branches in CoreFlow and SentenTree visualizations, and the number of sequences and events in Sequence Synoposis visualizations. The tutorial is included in the supplemental materials.


\vspace{-2mm} 
\subsubsection{Pre-Screening Test}
\vspace{-1.5mm}
\revise{We sampled the first 1000 events of a baseball game dataset (it was not a part of the main experiment) and created six questions to test the participant's literacy in terms of different low-level tasks.}
If a participant answers at least $50\%$ of the questions correctly, they are invited to the main experiment. The supplemental materials include the questions from the pre-screening test.
\vspace{-2mm}
\subsubsection{Main Experiment}
\label{sec:main_expr}
\vspace{-1.5mm}
In the main experiment, a brief description of the dataset and the problem domain is initially presented to the participant.
For example, the following text explains the emergency department dataset:

``\textit{The data is a sample of 100 patients, showing how a patient moves through the hospital over time.
All the possible events found in the data are: arrival at the hospital (Arrival), going to the emergency room (Emergency), to the ICU (ICU), to normal floor room (Floor) and discharged alive (Discharge-Alive) or not (Die).}''

In each of the following pages, the participant sees an insight and one visual summary generated by one of the three techniques. To eliminate any order effects,  we use the loop and merge feature on Qualtrics to randomize the order of techniques. The participant rates each technique on a $7$ point Likert scale, evaluating its ability to accurately match the insights for a given dataset. \revise{we conducted a preliminary study with ten participants to verify the soundness of our study method and estimate the average completion time.}

As described in Section \ref{sec:study_design}, each participant rates 
a total of nine (3 Insights $\times$ 3 techniques = $9$) visual summaries. 
They also provide a text justification for the ratings assigned. \revise{ In the pilot study, we asked} for a justification for each of the nine ratings, but the feedback from the participants indicated that this was too time consuming and the responses were very similar for the same technique. We thus only collect three justifications, one for each technique, from each participant. In addition to the ratings and text justifications, we also recorded the time they spent on each visual summary. 
\vspace{-2mm}


\subsection{Participants}\label{subsec: participants}
 \vspace{-1mm}
We recruited participants from multiple sources.   
We started recruiting participants with $\ge$ $90\%$ HIT Approval Rate on Mechanical Turk but encountered issues with the quality of responses; some participants did not pass the pre-screening test or did not understand the visual summaries- which was evident based on the text justification. Only $29.2\%$ of responses were high quality. Therefore, we switched to Prolific which has better filtering options. We recruited US residents who were at least $18$ years old, had an approval rate of $\ge$ $ 95\%$, and used a desktop. We limited educational status to those who attended technical/community college or had an undergraduate degree. We also recruited participants from the student body at our university. All of them had at least a Bachelor's degree. 
\revise{We finalized $180$ participants, yielding 1620 observations (180 Participants $\times$ 3 Techniques $\times$ 3 Insights). We made sure each condition has the same number of observations.}
All participants were compensated above the minimum age. All participants were compensated above minimum wage. Table \ref{tbl:participant} displays the acceptance rate for each platform.
 
 \begin{table}[th]\fontfamily{ptm}\selectfont\small
\centering
 {\small \begin{tabular}{cccc}
\arrayrulecolor{lightgray}
\hline		
\rowcolor[gray]{.95}  \multicolumn{1}{|c|}{\textbf{Source}} & \multicolumn{1}{|c|}{\textbf{Total participant}} & \multicolumn{1}{|c|}{\textbf{Accepted}}  & \multicolumn{1}{|c|}{\textbf{Acceptance Rate}}  \\  \hline  

MTurk	&	24	 & 7 & 29.2\%	\\ \hline	
Prolific	&	150	 & 135 & 90.0\%	\\ \hline	
Students	&	61	 & 55 & 90.2\%	\\ \hline	
\textbf{Total} &	\textbf{235}	 & \textbf{197} & \textbf{83.82\%}	\\ \hline
  \multicolumn{2}{c}{\textbf{Further Filtered}} & 17 &  \\ \hline
    \multicolumn{2}{c}{\textbf{Finally Accepted}} & \textbf{180} &   \\
\end{tabular}}
\setlength\belowcaptionskip{-6pt}
\setlength\abovecaptionskip{3
pt}
\caption{Participant details} \label{tbl:participant}
\end{table}

 \section{Analysis of Likert Scale Ratings of Visual Summaries}
 \label{sec:score_finings}


We first analyze the effects of various independent factors on summary quality, as measured using the Likert scale ratings assigned by the \revise{participants}. We perform exploratory analysis by creating visualizations of aggregated ratings, and 
build linear mixed-effects models to assess the effects of the independent variables. \revise{Mixed effects models are appropriate for our multi-level study design involving repeated measures, and have been used in similar empirical studies \cite{Liu_2014, Kim_2018}.}
 We model \textit{technique}, \textit{granularity}, and \textit{task} as fixed effects and \textit{participant}, \textit{dataset}, and \textit{insight} as random effects to extend the findings beyond the participants and datasets used in the study. Insights are nested under datasets in the models since each dataset has a unique set of insights.

We test the statistical significance of the fixed effects using likelihood-ratio tests \cite{winter2013linear}: we build a full model
(with the fixed effect in question) and a reduced model (without the
fixed effect in question), and compare these models to obtain p-values. \revise{The degrees of freedom for the comparison model is the difference in estimated parameters between the full and reduced model.}
The linear mixed-effects models are implemented using the R package lme4 \cite{bates2014fitting}. Below we report the results.
\vspace{-4mm}
\subsection{Technique Influences Visual Summary Quality}
\vspace{-1mm}
Figure \ref{fig:TDMdist} shows the distribution of ratings for each technique, as well as a more detailed breakdown of the ratings by task and dataset. Overall, Sequence Synposis (mean rating 3.86) outperforms SentenTree (mean 3.35) and CoreFlow (mean 2.95). 

This observed pattern in the figure is confirmed in the statistical analysis. We find a strong main effect of technique on the Likert scale ratings based on likelihood-ratio tests on random intercept models: ${\chi}^2 (2, N=1620) = 74.14$, $p<0.001$. 
For CoreFlow, the estimated intercept is ($3.03\pm $0.43 std. error). The estimated SentenTree intercept is $0.39$ higher ($3.42\pm 0.10$ std. error). The estimate for Sequence Synopsis is the highest at ($3.94 \pm 0.10$ std. error). 
The random intercept models assume that the effects of technique are the same for all the participants and all the datasets. Following the recommendations by Barr et al. ~\cite{BARR2013255},
we also build random slope models, the effect remains significant: ${\chi}^2 (2, N=1620) = 46.30$, $p<0.001$ where we assume the effects of technique vary for each participant, and ${\chi}^2 (2, N=1620) = 11.786$, $p<0.01$, where we assume the effects of technique vary for each dataset. 



We also built models for pairwise technique comparison \revise{, tested the significance and calculated the effect size in table \ref{tab:effectlikert}.} All the paired models show statistically significant effects of technique, implying that Sequence Synopsis significantly performs better than SentenTree, and SentenTree significantly outperforms CoreFlow. 

\begin{table}\fontfamily{ptm}\selectfont\small
\centering

 {\small \begin{tabular}{ p{0.15\linewidth} p{0.25\linewidth}  p{0.05\linewidth}  p{0.08\linewidth}}
\arrayrulecolor{lightgray}
\hline		
\rowcolor[gray]{.95} \multicolumn{1}{|c|}{\textbf{Technique1}} & \textbf{Technique2} & \multicolumn{1}{|c|}{\textbf{Rating}} & \multicolumn{1}{|c|}{\textbf{Time}} \\  \hline  

CoreFlow	& SentenTree	& $0.16$ & $0.09$ \\ \hline	
CoreFlow	& SequenceSynopsis		&	$0.33$ &	$0.36$ \\ \hline	
SentenTree	& SequenceSynopsis		&	$0.24$ 	&	$0.33$ \\ \hline	

\end{tabular}}
\setlength\belowcaptionskip{-14pt}
\setlength\abovecaptionskip{3
pt}
\caption{Cohen's d values for effect size estimation} \label{tab:effectlikert}
 \vspace{-2mm}
\end{table}



\begin{figure}[htbp]
\centering
\begin{subfigure}[b]{0.5\textwidth}
     \centering
     \includegraphics[width=\textwidth]{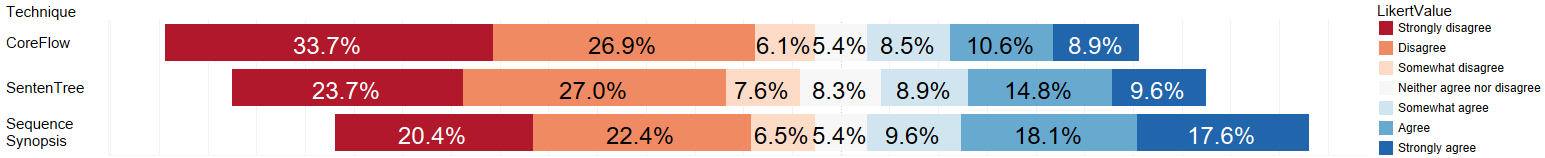}
     \caption{Likert scale ratings distribution across Technique}
     \label{fig:TDMdist_1}
 \end{subfigure}
 \begin{subfigure}[b]{0.5\textwidth}
     \centering
     \includegraphics[width=\textwidth]{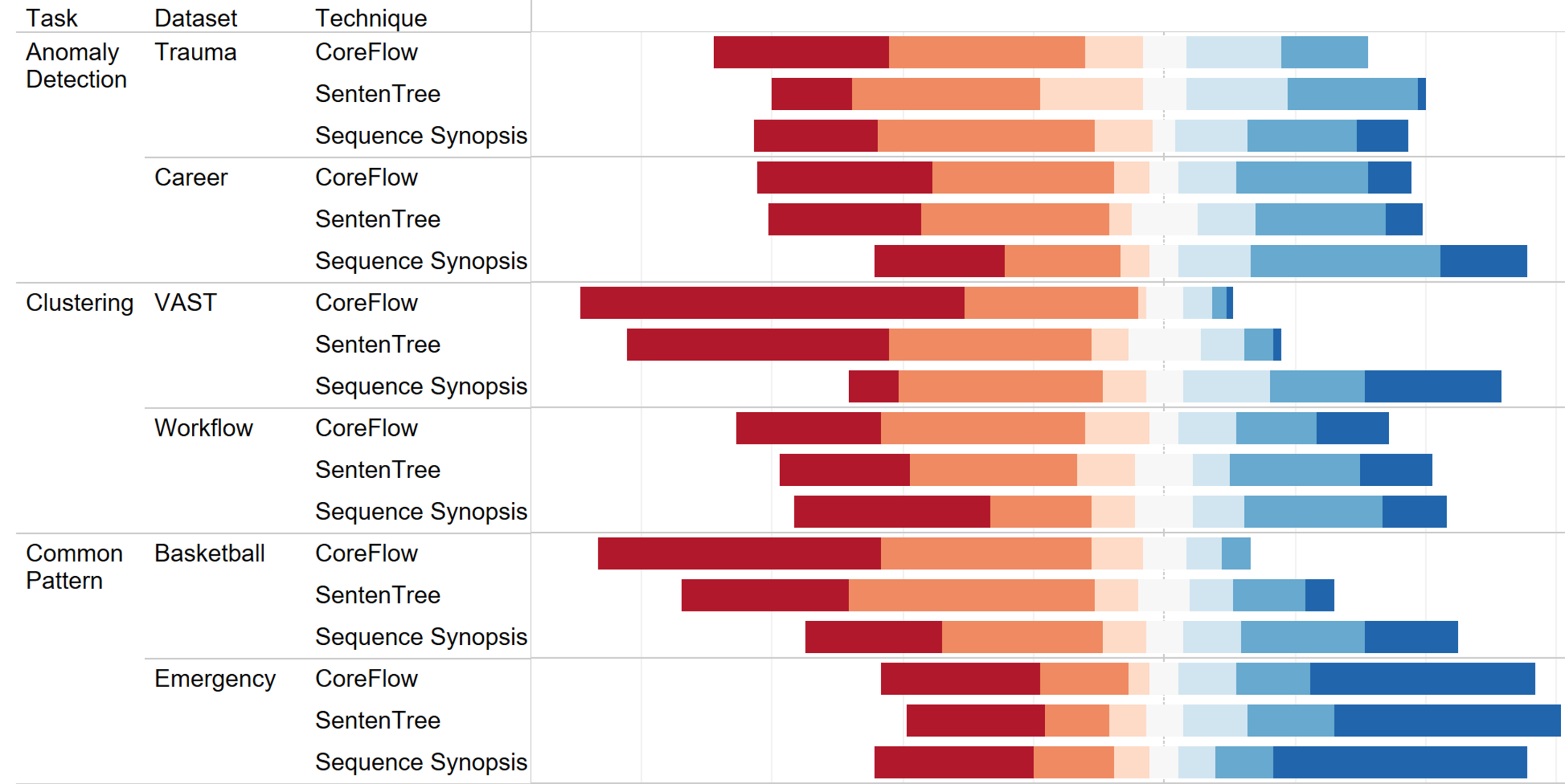}
     \caption{Likert scale ratings distribution across Task, Dataset and Technique}
     \label{fig:TDMdist_2}
 \end{subfigure}
  \setlength{\belowcaptionskip}{-8pt}
    \setlength{\abovecaptionskip}{0pt}
\caption{(a) the percentages of Likert scale ratings for each technique, out of the three techniques, Sequence Synopsis has the highest mean, followed by SentenTree and CoreFlow; (b) a detailed break-down of the ratings by task and dataset.} 
\label{fig:TDMdist}
\end{figure}








\vspace{-1mm}
\subsection{Granularity Matters for SentenTree}
\vspace{-1mm}
Based on likelihood ratio tests on random intercept models, we can not find statistically significant effect of granularity in the overall Likert scale ratings. 
We also analyze the within-group effect of granularity for all three techniques. The effect is not statistically significant for CoreFlow and Sequence Synopsis. However, granularity has a significant effect on the ratings for SentenTree in random intercept models: ${\chi}^2 (1, N=540) = 6.54$, $p<0.05$. 
The effect remains significant for random slope model: ${\chi}^2 (1, N=540) = 4.48$, $p<0.05$. Higher levels of abstraction in visual summaries produced by SentenTree lead to lower ratings. The rating decreases by $0.15$ for every 5\% increase in granularity.


We find significant interaction effects between technique and granularity: ${\chi}^2 (2, N=1620) = 16.354$, $p<0.001$. The effects of technique vary by granularity: 
\revise{for every 5\% increase in granularity, the rating drops by $0.006$ for CoreFlow  and by $0.15$ for SentenTree, but rises by 0.09 for Sequence Synposis.}

\vspace{-2mm}
\subsection{Task Has No Significant Effect on Rating}
\vspace{-1mm}

We do not find significant main effects for tasks on the Likert scale rating. 
We also model the interaction effects between \textit{task} and \textit{technique}, and between \textit{task} and \textit{granularity}. The technique-task interaction has a weak significance: ${\chi}^2 (4, N=1620) = 8.92$, $p<0.1$, but the granularity-task interaction is not significant.
For CoreFlow, the estimated intercept  for the Anomaly Detection task  is $3.15$, with the intercept for the Clustering task being $0.56$ lower at ($2.59 \pm 0.60$ std. error) and for the Common Pattern task is $0.12$ higher at ($3.27 \pm 0.60$  std. error). All three techniques have the highest rating for Common Pattern task, with Sequence Synopsis performing best (intercept estimate $4.02 \pm 0.26$  std. error). 
\begin{table}\fontfamily{ptm}\selectfont\small
\centering
 {\small \begin{tabular}{ccccc}
\arrayrulecolor{lightgray}
\hline		
\rowcolor[gray]{.95} \multicolumn{1}{|c|}{\textbf{Variables}} & \multicolumn{1}{|c|}{\textbf{Technique}} & \multicolumn{1}{|c|}{\textbf{Task}}   & \multicolumn{1}{|c|}{\textbf{Granularity}}  \\  \hline  

Likert Rating	&	\checkmark	 &  \ding{55} & \checkmark (SentenTree)	\\ \hline	
Completion Time	&	\checkmark	  & \ding{55} & \checkmark	\\ \hline	
\end{tabular}}
\setlength\belowcaptionskip{-15pt}
\setlength\abovecaptionskip{3
pt}
\caption{Analysis summary of statistical significance} \label{tab:significance}
\end{table}










\vspace{-1.5mm}
\section{Analysis of Completion Time}

Our analysis of the completion time – the time participants spent on evaluating each visual summary – is similar to our analysis of Likert scale ratings.
We use log-transformed completion time as the response variable to build linear mixed effects models and conduct likelihood ratio tests as mentioned in Section \ref{sec:score_finings}. Table \ref{tab:significance} shows our statistical analysis summary.




\vspace{-2mm}
\subsection{Technique Influences Completion Time}
\vspace{-1mm}

\begin{figure}[ht]
\centering
  \includegraphics[width=0.48\textwidth]{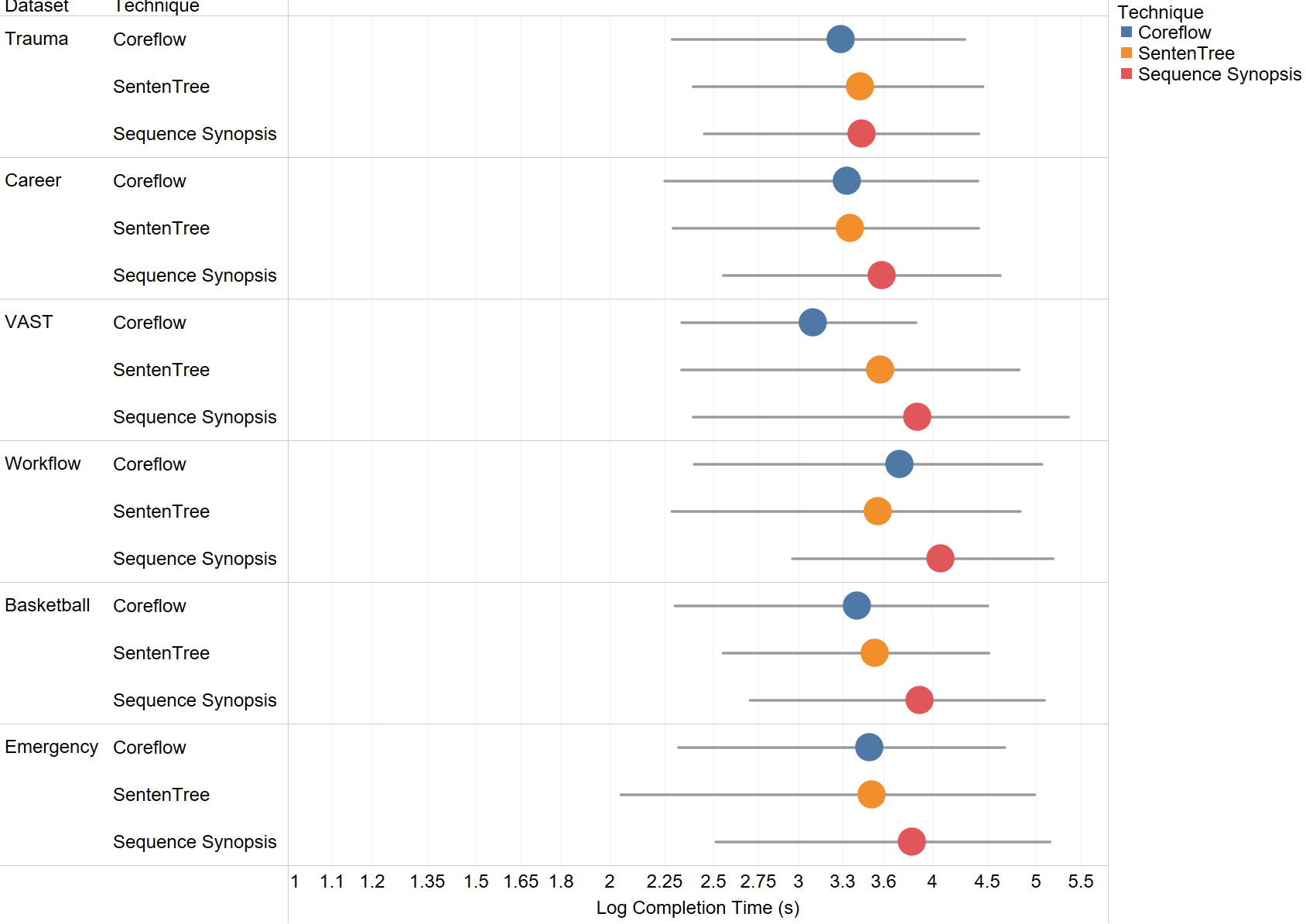}
  \setlength{\belowcaptionskip}{-6pt}
  \setlength{\abovecaptionskip}{-4pt}
  
  \caption{Bootstrapped means and 95\% confidence intervals for
log completion times across datasets and techniques. 
}
  \label{fig:CompletionTimeTech}
\end{figure}

To illustrate the effect sizes, we  compute the bootstrapped means and 95\% confidence intervals for log-transformed completion time by sampling individuals with replacement (Figure  \ref{fig:CompletionTimeTech}).  
Overall, participants spend more time evaluating visual summaries generated by Sequence Synposis compared to SentenTree and CoreFlow. 




The statistical findings corroborate the observed pattern. Based on likelihood-ratio tests on random intercept models, we find a significant main effect of technique on log completion time: 
${\chi}^2 (2, N=1620) = 98.92$, $p<0.001$. 
For CoreFlow, the estimated intercept is ($3.43\pm 0.14$ std. error). The estimated SentenTree intercept is $0.1$ higher ($3.53\pm 0.04$ std. error). The estimate for sequence synopsis is the highest ($3.82 \pm 0.04$ std. error). 
We also build random slope models, the effects remain significant: ${\chi}^2 (2, N=1620) = 52.66$, $p<0.001$, 
where we assume the effects of technique vary for each
participant, and ${\chi}^2 (2, N=1620) = 24.71$, $p<0.001$, where we assume
the effects of technique vary for each dataset.

Additionally, we developed models for pairwise technique comparison,  tested the significance \revise{ and calculated effect sizes (Table  \ref{tab:effectlikert})}.  All the paired models show statistical significance.

\vspace{-2mm}
\subsection{Task Has No Significant Effect on Completion Time}
\vspace{-1mm}



We find no main effects of task on completion time, nor significant interaction effects between task and granularity.
The 
task-technique interaction, however, is statistically significant:
${\chi}^2 (2, N=1620) = 15.77$, $p<0.01$. 
The projected intercept for the Anomaly Detection task in CoreFlow is $3.43$, with the intercepts for the Clustering and Common Pattern tasks being $0.25$ ($3.68 \pm 0.18$ std. error) and $0.21$ ($3.64 \pm 0.18$ std. error) higher, respectively.

The common pattern identification task has the greatest estimate for CoreFlow (estimate $3.44$).  The clustering task take the longest time for SentenTree (estimate $3.56$) and Sequence Synopsis (estimate $3.97$).
All tasks involving rating CoreFlow are finished by participants in the shortest amount of time. 

\vspace{-2mm}
\subsection{Granularity Affects Completion Time}
\vspace{-1mm}






We discover that the level of granularity has a statistically significant impact on the log-transformed completion time based on likelihood ratio tests on random intercept models: ${\chi}^2 (1, N=1620) = 13.25$, $p<0.001$. The intercept estimate for granularity is $0.08$ lower, indicating that with every $5\%$ increase in granularity, the log-transformed completion time drops by $0.08$. 

We also analyze the within-group effect of granularity for all  techniques. The effect is not statistically significant for Sequence Synopsis, but is significant for CoreFlow (${\chi}^2 (1, N=540) = 8.04$, $p<0.01$), and SentenTree (${\chi}^2 (1, N=540) = 14.58$, $p<0.001$).



\vspace{-1mm}
\section{Analysis of Qualitative Data}



In addition to the Likert scale ratings, we also asked the participants to provide text responses to justify their ratings. These qualitative responses can help us better understand the criteria and reasoning used by the participants to assess the quality of visual summaries. 


\begin{figure}[t]
\centering
  \includegraphics[width=0.5\textwidth]{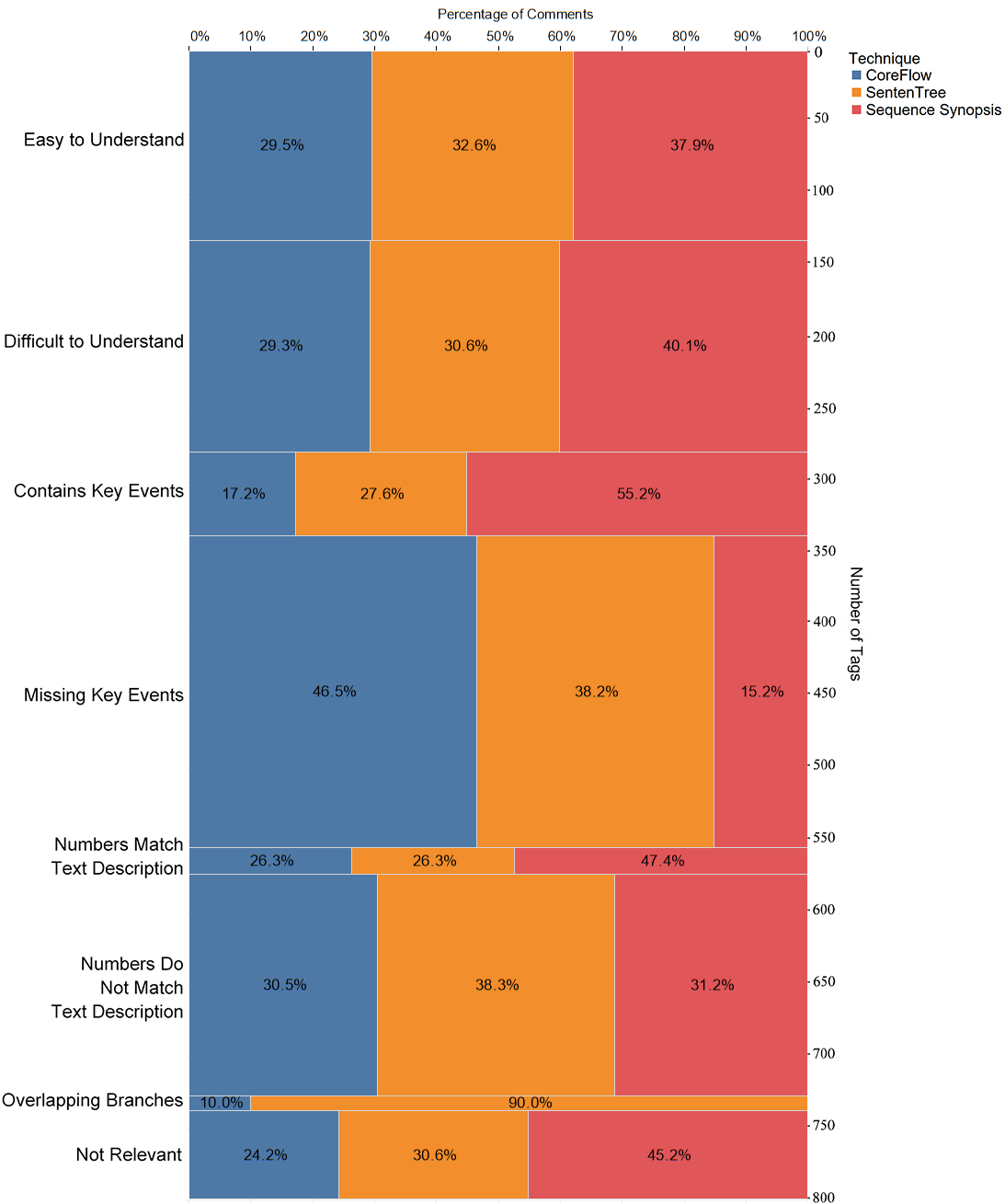}
  \setlength{\abovecaptionskip}{-3pt}
\setlength{\belowcaptionskip}{-13pt}
  \caption{Mosaic plot showing distribution of techniques for different tag categories. The bar height represents the number of comments for each tag category, and the bar width represents the percentage of comments contributed by a technique within each tag category. \textit{Missing Key Events} is the largest tag category. Most of the comments with this tag belongs to CoreFlow (46.5\%) }
  \label{fig:Tag}
  \vspace{-1.5mm}
\end{figure}



\vspace{-2mm}
We performed open coding on the responses from $180$ participants. 
Most comments touched upon two aspects of visual summary quality: \textit{content} (i.e., how closely the events and patterns included in the visualization match those mentioned in the insight), and \textit{interpretability} (i.e., how easy it is to read the visualization). We identified four tags related to content: 
\textit{Contains Key Events},
\textit{Missing Key Events},
\textit{Numbers Match Text Description}, and
\textit{Numbers Do Not Match Text Description}. The first two are concerned with whether important events are included in the visualization, and the remaining two focus on whether the quantitative information such as the number of sequences is consistent with the given insight. We also identified three tags related to interpretability: \textit{Easy to Understand},
\textit{Difficult to Understand}, and 
\textit{Overlapping Branches}, the first two are self-explanatory and the last one focuses on link crossing and cluttered views. 
Finally, we created a tag 
\textit{Not Relevant} to cover responses that are not intelligible or provide irrelevant information on the quality of visual summaries. 

We then manually labeled each response with these identified tags. A response can mention multiple aspects of the summary quality, hence assigned multiple tags. In total, we assigned 799 tags.
Figure \ref{fig:Tag} shows the number of comments associated with each tag, and the percentage distribution of each technique under each tag. The tag categories with the most comments are \textit{Missing Key Events} (217), \textit{Numbers Do Not Match Text Description} (154), \textit{Difficult to Understand} (147), and \textit{Easy to Understand} (132).
\vspace{-2mm}



\vspace{-2mm}
\subsection{Content}
\vspace{-1mm}
Many responses indicate that the visual summaries lack complete and consistent information compared to the given insights- this observation applies to all three techniques.
One participant remarks, ``\textit{Some of the activities in the facts were not shown in the image. The activities that were shown had numerical discrepancies to the fact}''.
This response falls under the \textit{Missing Key Events} and \textit{Numbers Do Not Match Text Description} tag. Similar comments were found across all three techniques.
On the other hand, some visualizations do conform to the insight: ``\textit{The numbers are all there, as far as my understanding is. That's why I chose strongly agree for each option.}'' 
The participants rate the image favorably when \textit{Numbers Match Text Description} and the image  \textit{Contains Key Events}.

Sequence Synopsis outperforms the other techniques in terms of including key events, with only 15.2\% of Missing Key Events tags associated with it, compared to 38.2\% for SentenTree and 46.5\% for CoreFlow. Furthermore, 55.2\% of \textit{Contains Key Events tags} are associated with Sequence Synopsis, compared to 27.6\% for SentenTree and 17.2\% for CoreFlow. CoreFlow only mines the most frequent event and trims subsequences preceding it, leading to the omission of less frequent events above the minimum support threshold. SentenTree iteratively extends the most common summary sequence, making it prone to missing events. In contrast, Sequence Synopsis uses a clustering and merging strategy that allows even less frequent events to form a pattern, preventing significant information loss during merging with more frequent events.


Sequence Synopsis also performs best in terms of numeric information accuracy. 
One interesting observation is techniques have almost equal share in the \textit{Numbers Do Not Match Text Description} tag. It might be due to how the algorithms group the sequences into patterns, or the fact that a pattern described in the insight is distributed across multiple branches or sequences. 
\vspace{-1mm}

\vspace{-2mm}
\subsection{Interpretability}

\vspace{-1mm}
Sequence Synopsis contributed the largest share of comments for both \textit{Easy to Understand} and \textit{Difficult to Understand}, followed by SentenTree, and CoreFlow. \revise{
The interpretability of the summaries varies by dataset and granularity. For the Basketball dataset at a granularity of 0.15, four out of five participants found Sequence Synopsis \textit{Difficult to Understand}, and none found it \textit{Easy to Understand}. On the contrary, for the Career dataset at a granularity of 0.3, three out of five participants found Sequence Synopsis \textit{Easy to Understand}, and none found it \textit{Difficult to Understand}.
}

Participants had mixed reactions to the branching patterns in CoreFlow and SentenTree visualizations. Some preferred CoreFlow's simplicity: ``\textit{This image is easy to understand and made the questions easy to answer as well because the path the numbers take is quite simple}'', while others found the single graph structure in SentenTree easier to comprehend than Sequence Synopsis: `` \textit{This image appears to be the same as the previous image, except with less information trees shown. Since there is only focus on one information tree in this image, I would say it is slightly easier to understand}''. One participant also noted the ease of following the branches to identify anomalies: `` \textit{This image was easy to read what the proper order of assessment was, what happened when there were deviations.  You could also determine the assessment outline after the deviations from the norm occurred.}''.

However, for some participants, the branching patterns were unfavorable: ``\textit{This image is not easy to follow because I'm not sure what one of the bifurcations means since it is not labeled. It also seems to be very confusing as to what were the sequence of events.}'' In finer granularity SentenTree representations, \textit{Overlapping Branches} are a prominent issue , which led to confusion and difficulty of understanding: ``\textit{The image is difficult to follow because there is a part of the graphic where pathways overlap and it is hard to attribute which data goes to which step}''.

The participants also have mixed reactions towards the individual linear sequence representation in Sequence Synopsis visualizations. Some participants prefer the distinction: ``\textit{It's easy to understand because it's step by step}''. Another comments: ``\textit{The images are easier to understand, because it delineates different threads of event \dots}''. On the other hand, 
some individuals find it difficult to consolidate information across sequences: ``\textit{It's tough to aggregate this information from the different vertical bars easily.}'' 


The patterns related to content can explain the Likert scale ratings. Sequence Synopsis may have received higher ratings because it 
includes more key events and accurate numeric information than the other algorithms.  The missing or inconsistent information may have contributed to 
CoreFlow and SentenTree obtaining a lower rating. We did not observe any strong correlation between accuracy and granularity, which is consistent with our finding in section \ref{sec:score_finings} that granularity has no impact on Likert scale ratings. 

In comparison with content, the relationship between interpretability and the ratings is less clear.
The technique with the highest Likert score ratings, Sequence Synopsis, accounts for the largest share (40.1\%) of comments with the \textit{Difficult to Understand} tag. The difficulty in understanding is in line with the fact that the participants spent more time evaluating the Sequence Synopsis graphics.
However, the largest share ($37.9\%$) of comments with the tag \textit{Easy to Understand} also belong to Sequence Synopsis. Further investigation is needed to understand how Sequence Synopsis' interpretability changes according to different datasets and granularity. 

\vspace{-1mm}
\section{Discussion}

\begin{figure}[t]
\centering
  \includegraphics[width=0.5\textwidth]{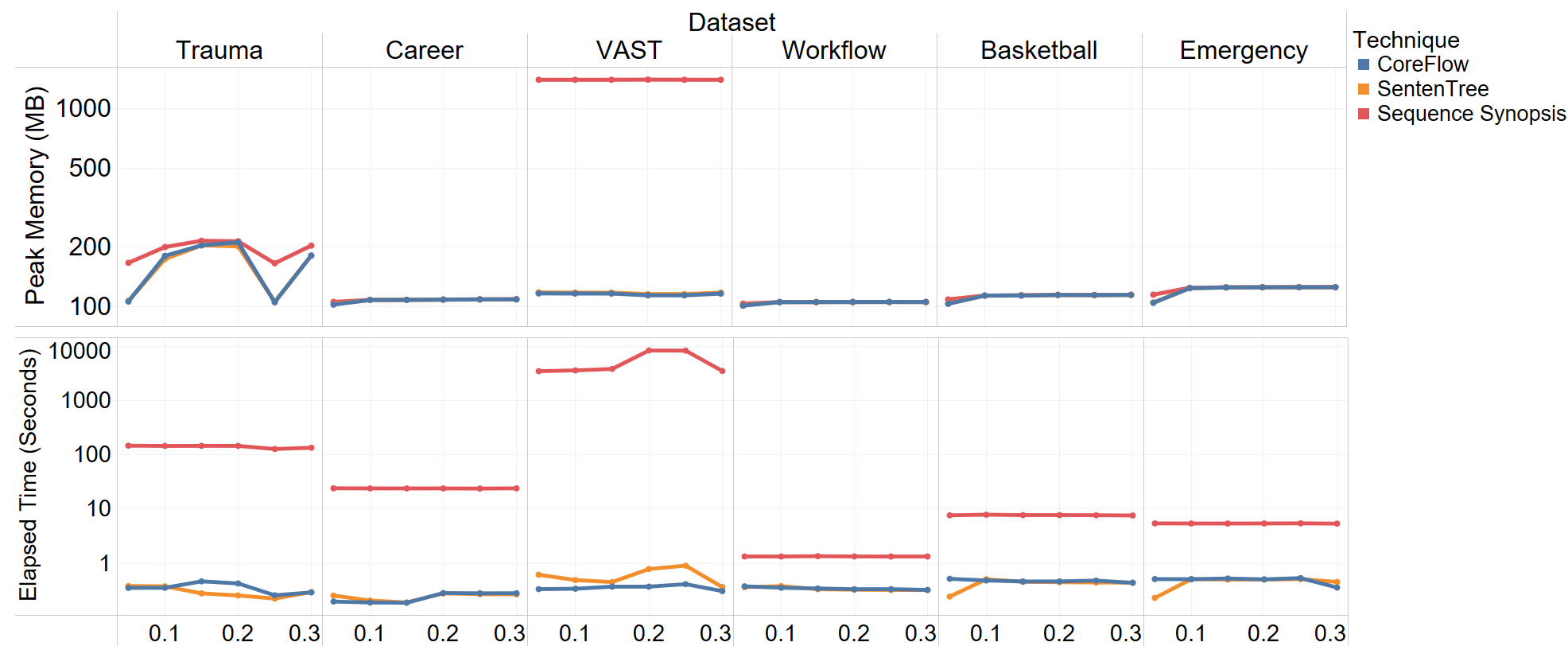}
  \setlength{\belowcaptionskip}{-10pt}
  \setlength{\abovecaptionskip}{-5pt}
  \caption{Time elapsed and  memory consumption across datasets at different granularities (finer to coarser, from left to right). The three techniques have similar memory consumption, except for the VAST dataset. Sequence Synopsis has the longest execution time.}
  \label{fig:Time}
  \vspace{-3mm}
\end{figure}


 \vspace{-1.5mm}

\bpstart{\revise{Possible Explanations of Rating Results}} Our analysis in section \ref{sec:score_finings} shows that Sequence Synopsis performs the best in terms of summary quality \revise{for all three tasks}.
\revise{Sequence Synopsis} uses a clustering and merging based information-theoretic mining approach with minimum description length principle. This enables the technique to penalize information loss while also taking   visual clutter reduction and number of summary sequences into account. These strategies help to produce accurate summary results. Both CoreFlow and SentenTree use frequent pattern based mining techniques, where the next most frequent event is added to the summary sequence in a greedy algorithmic approach.
There is no mechanism to account for information loss due to exclusion of less frequent events.
SentenTree has an option to regulate the overall number of events in the visualization,  but does not effectively control complexity or visual clutter in the branches, leading to lower Likert scale ratings. 


 \vspace{-1.5mm}
\bpstart{Trade-offs between Ratings and Reading/Computation Time} The task completion time is inversely correlated with technique ratings: while Sequence Synopsis performs best in terms of rating, it also requires the longest time for the participants to understand its visualization results. CoreFlow, on the other hand, obtains the lowest ratings,  due to its omission of key events, but produces simpler visual summaries that require the least comprehension time  among the three techniques. The need to strike a balance between summary complexity and accuracy likely applies generally to all visual summarization techniques for event sequences. 

\revise{Real-world applications must also consider computational time and memory cost.}
Figure \ref{fig:Time} displays the peak memory consumption and computational time of the techniques to mine the datasets at different levels of granularity (finer to coarser from left to right). Note that all the axis scales are logarithmic. 
\revise{Except for the Trauma and VAST datasets, the three techniques exhibit comparable maximum memory usage.}
Sequence Synopsis for the VAST dataset uses up to $1.4$ GB of memory at its maximum, significantly higher than the other techniques having maximum memory usage around $200$ MB.  
In terms of elapsed mining time, CoreFlow and SentenTree complete mining for each dataset under a minute, whereas Sequence Synopsis takes significant longer time, in particular, about one hour for the VAST dataset. 
If we take efficiency into account, Sequence Synopsis requires more computation time and memory resource in addition to human time. These factors are important when selecting techniques to use in practical situations.


 \vspace{-1.5mm}
\bpstart{Visual Summarization Techniques Need Improvement in General}
There is no silver bullet for creating perfect visual summaries of event sequence data, as demonstrated by our multifaceted analysis of human evaluation as well as computational resources.
Sequence Synopsis, the technique with highest average rating, has a score of only $3.86$ on a $7$ point Likert scale. There is still ample room for improvement in all the facets, including more accurate information content in the visual summaries,  less computational resources, and enhanced interpretability. 
\vspace{-2mm}


\vspace{-1mm}
\section{Limitations and Future Work}

\bpstart{Potential Difference from Original Implementation}
As described in \ref{sec: implementation}, we evaluated the techniques mentioned using our own implementations.
Although we follow the algorithm descriptions in the paper, our implementations might not accurately reflect the efficiency of the original code. 

 \vspace{-1.5mm}
\bpstart{\revise{Evaluating Interactive Systems}} 
\revise{Our current approach focuses on summary content and structure, which can be further refined and explored interactively to reach insights.
Future research can shed light on how the algorithms and interactive exploration together influence insight generation.
}

 \vspace{-1.5mm}
\bpstart{\revise{Inclusion of Larger Datasets}} 
\revise{The largest dataset in our experiment has $1000$ sequences. Real world datasets are often much larger in size and contains more event types.
This could potentially limit the generalizability of the findings to larger datasets. Despite this limitation, the experiment still provides valuable insights and serves as a starting point for further investigation.
} 

 \vspace{-1.5mm}
\bpstart{Interpreting and Controlling Granularity} 
\revise{SentenTree and CoreFlow use minimum support parameter to control granularity, where lower values indicate finer granularity and more events included in the visualization. Sequence Synopsis, on the other hand, regulates the number of summary sequences, rather than individual event frequency. The granularity variable in our experiments provides only an approximation and requires careful interpretation. Future research should explore ways to control granularity based on both pattern number and individual event frequency.}


 \vspace{-1.5mm}
\bpstart{Assessing Factors Influencing Technique Effectiveness} Our study evaluates the effectivenss of three techniques holistically. Each technique has multiple components working in conjunction. In particular, data reduction method (frequent pattern mining or information-theoretic clustering) and summary structure (linear sequences, tree, or graph) seem to play important roles. Further study is required to assess the effects of these individual components and their interaction on the outcome of visual summarization.

 \vspace{-1.5mm}
\bpstart{Guidelines for Selecting Visual Summarization Techniques} Our analysis can offer some crude guidance on choosing a technique for a given dataset and task. For instance, Sequence Synopsis is the best option when the data contains numerous important but sporadically occurring events.
CoreFlow may be the best option if the user is looking for a quick summary with the most common events.
SentenTree may be the best if the user is interested in the way events unfold in a branch-and-merge structure. However, our current analysis is insufficient to offer detailed and quantified technique recommendations. Further work is necessary to model the effects of dataset characteristics on technique effectiveness.  
 \vspace{-2mm}



\vspace{-2mm}
\section{Conclusion}
In this work, we present the experiment design and results comparing the effectiveness of three different visual summarization techniques for event sequence data. To the best of our knowledge, our study is the first of its kind to offer a comprehensive comparison of such techniques. The insight-based method can potentially inform future work that evaluates the effectiveness of new or existing techniques. Our quantitative and qualitative analysis results also provide insights on the techniques' performance, the trade-offs involved in event sequence visual summarization, and rooms of improvement for new summarization techniques.  

\bpstart{Acknowledgments}
We would like to thank Catherine Plaisant and Ben Shneiderman for providing access to the datasets and EventFlow. We also thank Hannah Bako and Yishan Ding for their assistance with the study design, as well as Esmotara Rima for helping with the front-end implementation.
\printbibliography
\end{document}